\documentclass[preprint,prc,aps,showpacs,showkeys,groupedaddress,floatfix]{revtex4}
\usepackage{graphics}
\usepackage{graphicx}
\usepackage{epsfig}
\usepackage{amssymb}
\usepackage{amsmath}

\begin{document}
\title{Asymptotic Iteration Method Solutions to the Relativistic Duffin-Kemmer-Petiau Equation}
\author{I. Boztosun, M. Karakoc, F. Yasuk and A. Durmus}
\affiliation{\it Department of Physics, Erciyes University, 38039
Kayseri, Turkey}
\begin{abstract}
A simple exact analytical solution of the relativistic
Duffin-Kemmer-Petiau equation within the framework of the asymptotic
iteration method is presented. Exact bound state energy eigenvalues
and corresponding eigenfunctions are determined for the relativistic
harmonic oscillator as well as the Coulomb potentials. As a
non-trivial example, the anharmonic oscillator is solved and the
energy eigenvalues are obtained within the perturbation theory using
the asymptotic iteration method.
\end{abstract}

 \maketitle
\section{Introduction}
 Exact analytical
solutions to relativistic wave equations are important in
relativistic quantum mechanics since the wave function contains all
the necessary information to describe a quantum system fully. There
are only a few potentials for which the relativistic Dirac,
Klein-Gordon and Duffin-Kemmer-Petiau (DKP) equations can be solved
analytically. So far, many methods such as the super-symmetric
(SUSY) \cite{1}, shape invariance \cite{2,3}, factorization and path
integral \cite{4,5,6,7} \emph{etc} have been developed to solve the
relativistic wave equations exactly, or quasi-exactly, for
potentials like Coulomb, harmonic oscillator, P\"{o}sch Teller and
exponential type ones. In recent years, an asymptotic iteration
method for solving second order homogeneous linear differential
equations has been proposed by Ciftci \emph{et al.} \cite{8,9,10}.
This method has been applied to solve the non-relativistic radial
Schr\"{o}dinger and Dirac equations for various potentials
\cite{10}.

Since the DKP equation is being increasingly used to describe the
interactions of relativistic spin-0 and spin-1 bosons
\cite{11,12,13,14,15,16,17,18,19}, it would be interesting to probe
whether the DKP equation is amenable to exact solutions in the
framework of the asymptotic iteration method (AIM). This is
precisely the aim of this paper.

In the next section, we explain the AIM briefly and show how to
solve a second-order homogeneous differential equation. Then, we
introduce the DKP oscillator and Coulomb problems and obtain their
exact eigenvalues and eigenfunctions. In section \ref{anharmonic},
we present the solution of the anharmonic oscillator as a nontrivial
example within the perturbation theory. Finally, in the last
section, we provide our summary and conclusion.

\section{Basic Equations of the Asymptotic Iteration Method
(AIM)}\label{aim} We briefly outline the asymptotic iteration method
here; the details can be found in references \cite{8,9,10}. The
asymptotic iteration method was proposed to solve second-order
differential equations of the form
\begin{equation}\label{diff}
  y''=\lambda_{0}(x)y'+s_{0}(x)y
\end{equation}

where $\lambda_{0}(x)\neq 0$ and s$_{0}$(x), $\lambda_{0}$(x) are in
C$_{\infty}$(a,b). The variables, s$_{0}$(x) and $\lambda_{0}$(x),
are sufficiently differentiable. The differential equation
(\ref{diff}) has a general solution \cite{8}

\begin{equation}\label{generalsolution}
  y(x)=exp \left( - \int^{x} \alpha dx^{'}\right ) \left [C_{2}+C_{1}
  \int^{x}exp  \left( \int^{x^{'}} \lambda_{0}(x^{''})+2\alpha(x^{''}) dx^{''} \right ) dx^{'} \right
  ]
\end{equation}

if $n>0$, for sufficiently large $n$,

\begin{equation}\label{quantization}
\frac{s_{n}}{\lambda_{n}}=\frac{s_{n-1}}{\lambda_{n-1}}=\alpha
\end{equation}
where

\begin{equation}\label{iter}
  \lambda_{n}=\lambda_{n}'+s_{n-1}+\lambda_{0}\lambda_{n-1}\hspace{1cm} \mbox{and} \hspace{1cm}
  s_{n}=s_{n-1}'+s_{0}\lambda_{n-1}
\end{equation}

The quantization condition of the method together with equation
(\ref{iter}) can also be written as follows

\begin{equation}\label{kuantization}
  \delta(x)=\lambda_{n+1}(x)s_{n}(x)-\lambda_{n}(x)s_{n+1}(x)=0
\end{equation}

For a given potential, the idea is to convert the relativistic wave
equation to the form of equation (\ref{diff}). Then, s$_{0}$ and
$\lambda_{0}$ are determined and s$_{n}$ and $\lambda_{n}$
parameters are calculated. The energy eigenvalues are then obtained
by the condition given by equation (\ref{kuantization}). However,
the wave functions are determined by using the wave function
generator, namely $exp \left( - \int^{x} \alpha dx^{'}\right )$.

In this study, we seek the exact solution of DKP equation for which
the relevant second order homogenous linear differential equation
takes the following general form,
\begin{equation}
 {y}'' = 2\left( {\frac{ax^{N + 1}}{1 - bx^{N + 2}} -
\frac{\left( {m + 1} \right)}{x}} \right){y}' - \frac{wx^N}{1 -
bx^{N + 2}}y
\end{equation}
If this equation is compared to equation (\ref{diff}), it entails
the following expressions
\begin{equation}\label{snln}
\lambda _0 = 2\left( {\frac{ax^{N + 1}}{1 - bx^{N + 2}} -
\frac{\left( {m + 1} \right)}{x}} \right)   \hspace{1cm}  s_0 (x) =
- \frac{wx^N}{1 - bx^{N + 2}}
\end{equation}
while the condition (\ref{quantization}) yields for
$N$=-1,0,1,2,3,....
\begin{eqnarray}
w_n^m (-1) & = & n\left( {2a + 2bm + (n + 1)b} \right)\\
w_n^m (0) & = & 2n\left( {2a + 2bm + (2n + 1)b} \right) \\
w_n^m (1) & = & 3n\left( {2a + 2bm + (3n + 1)b} \right) \\
w_n^m (2) & = & 4n\left( {2a + 2bm + (4n + 1)b} \right) \\
w_n^m (3) & = & 5n\left( {2a + 2bm + (5n + 1)b} \right) \\
\ldots \emph{etc} \nonumber
\end{eqnarray}
Hence, these formulae are easily generalized as;
\begin{equation}
w_n^m (N) = b\left( {N + 2} \right)^2n\left( {n + \frac{\left( {2m
+ 1} \right)b + 2a}{\left( {N + 2} \right)b}} \right)
\end{equation}
The exact eigenfunctions can be derived from the following
generator:
\begin{equation}\label{ef}
y_n (x) = C_2 \exp \left( { - \int\limits^x {\alpha _k dx^{'}} }
\right)
\end{equation}
Using equation (\ref{quantization}) and equation (\ref{snln}), the
eigenfunctions are obtained as follows;
\begin{small}
\begin{eqnarray*}
y_0 (x) & = & 1 \\
y_1 (x) & = & - C_2 (N + 2)\sigma \left( {1-\frac{b\left( {\rho + 1}\right)}{\sigma }x^{N + 2}} \right) \\
\\
y_2 (x) & = & C_2 (N + 2)^2\sigma \left( {\sigma + 1} \right)\left(
{1 - \frac{2b\left( {\rho + 2} \right)}{\sigma }x^{N + 2} +
\frac{b^2\left( {\rho + 2} \right)\left( {\rho + 3} \right)}{\sigma
\left( {\sigma + 1} \right)}x^{2(N + 2)}} \right)
\\
 y_3(x) & = & - C_2\frac{\sigma\left({\sigma+1}\right) \left({\sigma+2}
\right)}{\left({N+2}\right)^{-3}} \nonumber \\  & \times & \left
({1-\frac{3b\left( {\rho + 3}\right)} \sigma x^{N+2}+
\frac{3b^2\left({\rho+3}\right)\left(
{\rho+4}\right)}{\sigma\left({\sigma+1} \right)}x^{2(N+2)}
-\frac{b^3\left({\rho+3}\right)\left({\rho+4}\right)\left({\rho+ 5}
\right)}{\rho \left( {\rho + 1} \right)\left( {\rho + 2}
\right)}x^{3\left( {N + 2} \right)}} \right ) \nonumber \\
\ldots \emph{etc}
\end{eqnarray*}
\end{small}
Finally, the following general formula for the exact solutions
$y_n(x)$ is acquired as;
\begin{equation}\label{efson}
y_n (x) = \left( { - 1} \right)^nC_2 (N + 2)^n\left( \sigma
\right)_n { }_2F_1 ( - n,\rho + n;\sigma ;bx^{N + 2})
\end{equation}

where $(\sigma )_n $=$\frac{\Gamma ( {\sigma + n} )}{\Gamma (\sigma
}, \quad \sigma$ = $\frac{2m + N + 3}{N + 2}$ \quad \mbox{and} \quad
$\rho$ = $\frac{( {2m + 1} )b + 2a}{( {N + 2} )b}$.


%
\section{DKP Harmonic Oscillator}
In this section, the Duffin-Kemmer-Petiau formalism \cite{12,13} is
briefly sketched and the DKP oscillator is solved using AIM.
Generally, the first order relativistic Duffin-Kemmer-Petiau
equation for a free spin zero or spin one particle of mass m is

\begin{equation}
\label{eq1} ( {c\mathbf{\beta}.\textbf{p}+mc^2})\psi=i\hbar\beta^0
\frac {d\psi}{dt}
\end{equation}
where $\beta ^{\mu }$ ($\mu $= 0, 1, 2, 3) matrices satisfy the
commutation relation
\begin{equation}
\label{eq2} \beta ^\mu \beta ^\nu \beta ^\lambda +\beta ^\lambda
\beta ^\nu \beta ^\mu =g^{\mu \nu }\beta ^\lambda +g^{\nu \lambda
}\beta ^\mu
\end{equation}
which defines the so-called Duffin-Kemmer-Petiau (DKP) algebra. The
algebra generated by the four $\beta$ matrices has three irreducible
representations: a ten dimensional one that is related to S=1, a
five dimensional one relevant for S=0 (spinless particles) and a one
dimensional one which is trivial.

In the spin-0 representation, $\beta ^\mu $ are $5\times 5$
matrices defined as ($i = 1, 2, 3$)
\begin{equation}
\label{eq3} \beta ^0=\left( {{\begin{array}{*{20}c}
 \theta \hfill & {\tilde {0}} \hfill \\
 {\bar {0}_T } \hfill & \textbf{0} \hfill \\
\end{array} }} \right),
\quad \beta ^i=\left( {{\begin{array}{*{20}c}
 {\tilde {0}} \hfill & {\rho ^i} \hfill \\
 {-\rho _T^i } \hfill & \textbf{0} \hfill \\
\end{array} }} \right)
\end{equation}
with $\tilde {0}$, $\bar {0}$, $\textbf{0}$ as $2\times 2$,
$2\times 3$, $3\times 3$ zero matrices, respectively, and
\begin{equation}
\label{eq4} \theta=\left( {{\begin{array}{*{20}c}
 0 \hfill & 1 \hfill \\
 1 \hfill & 0 \hfill \\
\end{array} }} \right),
\quad \rho^1=\left( {{\begin{array}{*{20}c}
 -1 \hfill & 0 \hfill & 0 \hfill\\
 0 \hfill & 0 \hfill & 0 \hfill\\
\end{array} }} \right),
\quad \rho^2=\left( {{\begin{array}{*{20}c}
 0 \hfill & -1 \hfill & 0 \hfill\\
 0 \hfill & 0 \hfill & 0 \hfill\\
\end{array} }} \right),
\quad \rho^3=\left( {{\begin{array}{*{20}c}
 0 \hfill & 0 \hfill & -1 \hfill\\
 0 \hfill & 0 \hfill & 0 \hfill\\
\end{array} }} \right)
\end{equation}
For spin one particles, $\beta^\mu$ are $10\times 10$ matrices
given by
\begin{equation}
\label{eq5} \beta^0=\left( {{\begin{array}{*{20}c}
 0 \hfill & {\bar 0} \hfill & {\bar 0} \hfill & {\bar 0} \hfill\\
 {\bar 0}^T \hfill & \textbf {0} \hfill  & \textbf {I} \hfill & \textbf {I}
\hfill\\
 {\bar 0}^T \hfill & \textbf {I} \hfill & \textbf {0} \hfill & \textbf {0}
\hfill\\
{\bar 0}^T \hfill & \textbf {0} \hfill & \textbf {0} \hfill &
\textbf {0}
\hfill\\
\end{array} }} \right),
\quad \beta^i=\left( {{\begin{array}{*{20}c}
 0 \hfill & {\bar 0} \hfill & e_i \hfill & {\bar 0} \hfill\\
 {\bar 0}^T \hfill & \textbf {0} \hfill  & \textbf {0} \hfill & -is_i \hfill\\
 e_i^T \hfill & \textbf {0} \hfill & \textbf {0} \hfill & 0 \hfill\\
{\bar 0}^T \hfill & -is_i \hfill & \textbf {0} \hfill & 0 \hfill\\
\end{array} }} \right)
\end{equation}
where $s_{i}$ are the usual $3\times 3$ spin one matrices
\begin{equation}
\label{eq6} {\bar 0}=\left( {{\begin{array}{*{20}c}
 0 \hfill & 0 \hfill & 0 \hfill\\
\end{array} }} \right),
\quad e_1=\left( {{\begin{array}{*{20}c}
 1 \hfill & 0 \hfill & 0 \hfill\\
 \end{array} }} \right),
\quad e_2=\left( {{\begin{array}{*{20}c}
 0 \hfill & 1 \hfill & 0 \hfill\\
 \end{array} }} \right),
\quad e_3=\left( {{\begin{array}{*{20}c}
 0 \hfill & 0 \hfill & 1 \hfill\\
 \end{array} }} \right).
\end{equation}
$\textbf{I}$ and $\textbf{0}$ are the identity and zero matrices,
respectively. While the dynamical state $\psi_{DKP}$ is a five
component spinor for spin zero particles, it has ten component
spinors for $S=1$ particles.

For the external potential introduced with the non-minimal
substitution
\begin{equation}
\textbf{p} \to \textbf{p} - im\omega \eta ^0\textbf{r}
\end{equation}
where $\omega$ is the oscillator frequency and $\eta ^0$ = $2\beta
^{0^2}$ - 1, the DKP equation for the system is
\begin{equation}\label{dkpspinor}
\left[ {c\beta .(\textbf{p} - im\omega \eta ^0\textbf{r}) + mc^2}
\right]\psi = i\hbar \beta ^0\frac{d\psi }{dt}
\end{equation}
In the spin zero representation, the five component DKP spinor

\begin{equation}
\label{eq8} \psi(\textbf {r})=\left( {{\begin{array}{*{20}c}
\psi_{upper} \hfill\\
i\psi_{lower} \hfill\\
\end{array} }} \right)
\quad \mbox{with}~~~\psi_{upper}\equiv \left(
{{\begin{array}{*{20}c}
\phi \hfill\\
\varphi \hfill\\
\end{array} }} \right)
\quad \mbox{and}~~~\psi_{lower}\equiv \left( {{\begin{array}{*{20}c}
A_1 \hfill\\
A_2 \hfill\\
A_3 \hfill\\
 \end{array} }} \right).
\end{equation}
so that for stationary states the DKP equation can be written as
\begin{equation}
\begin{array}{l}
 mc^2\phi = E\varphi \, + ic(\textbf{p} + im\omega \textbf{r}).\textbf{A} \\
 mc^2\varphi = E\phi \\
 mc^2\textbf{A} = ic(\textbf{p} - im\omega \textbf{r})\phi \\
 \end{array}
\end{equation}

where $\textbf{A}$ is the vector $(A_1,A_2, A_3)$.\\
The five-component wavefunction $\psi$ is simultaneously an
eigenfunction of $J^2$ and $J_3$
\begin{equation}
J^2\left( {{\begin{array}{*{20}c}
\psi_{upper} \hfill\\
\psi_{lower} \hfill\\
\end{array} }} \right)=\left( {{\begin{array}{*{20}c}
L^2\psi_{upper} \hfill\\
(L+S)^2\psi_{lower} \hfill\\
\end{array} }} \right)=J(J+1)\left( {{\begin{array}{*{20}c}
\psi_{upper} \hfill\\
\psi_{lower} \hfill\\
\end{array} }} \right)
\end{equation}
\begin{equation}
J_3\left( {{\begin{array}{*{20}c}
\psi_{upper} \hfill\\
\psi_{lower} \hfill\\
\end{array} }} \right)=\left( {{\begin{array}{*{20}c}
L_3\psi_{upper} \hfill\\
(L_3+s_3)\psi_{lower} \hfill\\
\end{array} }} \right)=M\left( {{\begin{array}{*{20}c}
\psi_{upper} \hfill\\
\psi_{lower} \hfill\\
\end{array} }} \right)
\end{equation}
where the total angular momentum $J=L+S$ which commutes with
$\beta^0$, is a constant of the motion.


For $S=0$ DKP oscillator eigenstates problem, the most general
solution for a central problem \cite{13} is presented as follows

\begin{equation}
\label{eq12} \psi_{JM}(r)=\left( {{\begin{array}{*{20}c}
F_{nJ}(r)Y_{JM}(\Omega) \hfill\\
G_{nJ}(r)Y_{JM}(\Omega) \hfill\\
i\sum_LH_{nJL}(r)Y_{JL1}^M(\Omega) \hfill\\
\end{array} }} \right)
\end{equation}
where
\begin{equation}
\alpha_J = \sqrt {\left( {J + 1} \right) / \left( {2J + 1} \right)},
\quad \zeta_J = \sqrt {J / \left( {2J + 1} \right)}
\end{equation}

\begin{equation}
\begin{array}{l}
 F_{nJ} (r) = F(r),\quad G_{nJ} = G(r),\quad H_{n,J,J\pm
1} (r) = H_{\pm 1} (r) \\
\end {array}
\end{equation}
$\psi_{JM}$ of parity $(-1)^J$ is inserted into equation
(\ref{dkpspinor}) and the following equations are found.
\begin{equation}
\label{h1}
EF = mc^2G \\
\end{equation}
\begin{equation}\label{h2}
\hbar c\left( {\frac{d}{dr} - \frac{J + 1}{r} + \frac{m\omega
r}{\hbar }}
\right)F = - \frac{1}{\alpha _J }mc^2H_1  \\
\end{equation}
\begin{equation}\label{h3}
 \hbar c\left( {\frac{d}{dr} - \frac{J}{r} + \frac{m\omega r}{\hbar }}
\right)F = - \frac{1}{\zeta _J }mc^2H_{ - 1}
\end{equation}
\begin{equation}\label{h4}
- \alpha _J \left(
{\frac{d}{dr} + \frac{J + 1}{r} - \frac{m\omega r}{\hbar }} \right)H_1 + \\
 \zeta _J \left( {\frac{d}{dr} - \frac{J}{r} - \frac{m\omega r}{\hbar }}
\right)H_{ - 1} = \frac{1}{\hbar c}\left( {mc^2F - EG} \right)  \\
\end{equation}
From the above equations, if equations (\ref{h1}) to (\ref{h3}) are
inserted into equation (\ref{h4}), the homogenous second order
differential equation for the DKP harmonic oscillator \cite{13} is
obtained as;

\begin{equation}\label{hodiff}
\left( {\frac{d^2}{dr^2} + \frac{\left( {E^2 - m^2c^4}
\right)}{\left( {\hbar c} \right)^2} + \frac{3m\omega }{\hbar } -
\frac{m^2\omega ^2r^2}{\hbar ^2} - \frac{J\left( {J + 1}
\right)}{r^2}} \right)F(r) = 0
\end{equation}
If we define $E_{eff}$ = $\frac{\left( {E^2 - m^2c^4}
\right)}{\left( {\hbar c} \right)^2} + \frac{3m\omega }{\hbar }$ and
$k =\frac{m\omega }{\hbar }$, equation (\ref{hodiff}) becomes


\begin{equation}\label{hodiffarranged}
\left( {\frac{d^2}{dr^2} + E_{eff} - k^2r^2 - \frac{J\left( {J + 1}
\right)}{r^2}} \right)F(r) = 0
\end{equation}
The asymptotic iteration method requires selecting the wave function
as follows
\begin{equation}
F(r) = r^{J + 1}e^{ - \frac{1}{2}kr^2}f(r)
\end{equation}
equating it into equation (\ref{hodiffarranged}) leads to
\begin{equation}
\frac{d^2f(r)}{dr^2} -2\left( {kr - \frac{J + 1}{r}}
\right)\frac{df(r)}{dr} + \left( {E_{eff} - 3k - 2kJ} \right)f(r)=0
\end{equation}
where $\lambda _0$ = $2\left( {kr - \frac{J + 1}{r}} \right)$ and
$s_0 = 3k + 2kJ - E_{eff}$. By means of equation (\ref{iter}), we
may calculate $\lambda_n(r)$ and $s_n(r)$. This gives:
\begin{eqnarray}\label{sl}
\lambda _0 & = & 2\left( {kr - \frac{J + 1}{r}} \right) \nonumber \\
 s_0 & = & 3k + 2kJ - E_{eff}
 \nonumber \\
\lambda _1 & = & 5k + 2\frac{J + 1}{r^2} + 2kJ - E_{eff} + 4\left(
{kr
-\frac{J + 1}{r}} \right)^2 \nonumber \\
s_1 & = & 2\left( {3k + 2kJ - E_{eff} } \right)\left( {kr - \frac{J
+ 1}{r}}\right)
 \nonumber \\
\lambda _2 & = & - 4\frac{J + 1}{r^3} + 2\left( {kr - \frac{J +
1}{r}} \right)\left[ {4\left( {k + \frac{J + 1}{r^2}} \right) +
\left( {3k - 2kJ -
E_{eff} } \right) + 2\left( {kr - \frac{J + 1}{r}} \right)} \right] \nonumber \\
s_2 & = & \left( {3k + 2kJ - E_{eff} } \right)\left[ {7k + 4\frac{J
+ 1}{r^2}+2kJ - E_{eff} + 4\left( {kr - \frac{J + 1}{r}} \right)^2}
\right] \\
\ldots \emph{etc} \nonumber
\end{eqnarray}

Combining these results with the quantization condition given by
equation (\ref{kuantization}) yields
\begin{eqnarray}
 \frac{s_0 }{\lambda _0 } = \frac{s_1 }{\lambda _1 }\,\,\,\,\,\, \Rightarrow
\,\,\,\,\,\,\left( {E_{eff} } \right)_0 = 3k + 2kJ \\
 \frac{s_1 }{\lambda _1 } = \frac{s_2 }{\lambda _2 }\,\,\,\,\,\, \Rightarrow
\,\,\,\,\,\,\left( {E_{eff} } \right)_1 = 7k + 2kJ \\
 \frac{s_2 }{\lambda _2 } = \frac{s_3 }{\lambda _3 }\,\,\,\,\,\, \Rightarrow
\,\,\,\,\,\,\left( {E_{eff} } \right)_2 = 11k + 2kJ \\
\ldots \emph{etc} \nonumber
 \end{eqnarray}

When the above expressions are generalized, the DKP oscillator
eigenvalues turn out as
\begin{equation}\label{eigeneff}
(E_{eff})_n = k\left( {4n + 3 + 2J} \right)
\end{equation}
If one inserts the values of $k$ and $E_{eff}$ into equation
(\ref{eigeneff}), the relativistic energy spectrum of DKP oscillator
becomes
\begin{equation}
\frac{1}{2mc^2}\left( {E_{NJ}^2 - m^2c^4} \right) = N\hbar \omega
\end{equation}
where $N$ is the principal quantum number defined as $N=2n+J$. Our
result is in agreement with the result of reference \cite{13} for
the same potential.

As indicated in Section \ref{aim}, we can construct the
corresponding eigenfunctions by using the wave function generator
given by equation (\ref{ef}) and equation (\ref{sl}) where we obtain
$\lambda$ and $s$ values. Therefore, similar to equation
(\ref{efson}), the wave function $f_n (r)$ can be written:
\begin{equation} f_n (r) = \left( { - 1} \right)^nC_2
2^n(\sigma)_n~{ }_1F_1 \left( { - n,\sigma ;kr^2} \right)
\end{equation}
$F(r)$ ensues right away in the following form:
\begin{equation}
F(r) = r^{J + 1}e^{ - \frac{1}{2}kr^2} \left [\left( { - 1}
\right)^nC_2 2^n(\sigma)_n~{ }_1F_1 \left( { - n,\sigma
;kr^2}\right)\right]
\end{equation}
where
\begin{equation}
 \sigma = \frac{2J + 3}{2} \hspace{0.5cm} \mbox{and}  \hspace{0.5cm}
 \left( \sigma \right)_n = \frac{\Gamma \left( {\sigma + n} \right)}{\Gamma
\left( \sigma \right)}
\end{equation}

Using the wave function  $F(r)$, the wave functions $G(r)$,
$H_{1}(r)$ and $H_{ - 1}(r)$ can be easily obtained by using
equations (\ref{h1}) to (\ref{h4}).
\section{DKP Coulomb Potential}
We now apply the AIM method to the bound state problem of a spinless
charged pion ($\pi^{-}$) in the Coulomb field of a nucleus. If we
use following \emph{ansatz}:
\begin{equation}
a_{\pm} = \frac{mc^2 \pm E}{ \hbar c}, \quad \gamma = \alpha Z,
\quad \lambda_{\pi} = \frac{\hbar}{mc}, \quad\kappa = \frac{2}{
\hbar c} \sqrt{m^2c^4 - E^2}, \quad \xi= \frac {2\gamma E}{\kappa
\hbar c}, \quad \rho = \kappa r \label{kýsaltmalar}
\end{equation}
the system of coupled equations  for the Coulomb potential  becomes
\begin{equation}
\alpha_J \left(  \frac{d F}{ d\rho} -  \frac{J + 1}{ \rho} F \right)
 =  - \frac{1}{\kappa \lambda_{\pi} } H_1 \label{c1}
\end{equation}
\begin{equation}
\zeta_J \left(\frac{d F}{d\rho} + \frac{J}{\rho} F \right)  =
\frac{1}{\kappa\lambda_{\pi}} H_{-1} \label{c2}
\end{equation}
\begin{eqnarray}
- \alpha_J \left( \frac{d H_{1}}{d\rho} + \frac{J + 1}{\rho} H_{1}
\right)  + \zeta_J \left( \frac{d H_{-1}}{d\rho} - \frac{J}{\rho}
H_{-1} \right) \nonumber \\
 =  \kappa\lambda_{\pi}
  \left( \frac{a_+}{ \kappa} + \frac{\gamma}{\rho} \right)
  \left( \frac{a_-}{\kappa} - \frac{\gamma}{\rho} \right) F  \label{c3}
\end{eqnarray}

Eliminating $H_{1}$ and $H_{-1}$ in favor of $F$, the second-order
differential equation for the Coulomb potential becomes
\begin{equation}
{\frac {d^{2}F(\rho)}{d{\rho}^{2}}}+\left ({\frac {\xi}{\rho}}-\frac{1}{4}-
{\frac {J\left (J+1\right )-\gamma^2}{{\rho}^{2}}}\right )F(\rho)=0 \label{coulombdkp}
\end{equation}
Let the radial wave function be factorized as:
\begin{equation}
F(\rho)={\rho}^{\Lambda+1}{e^{-\frac{1}{2}\rho}}f(\rho)
\end{equation}
where
\begin{equation}
\Lambda=-\frac{1}{2}+\sqrt {(J+\frac{1}{2})^2-\gamma^{2}}
\end{equation}
Equation (\ref{coulombdkp}) becomes

\begin{equation}
{\frac {d^{2}f(\rho)}{d{\rho}^{2}}}-{\frac {\left (\rho-2\,
\Lambda-2\right )}{\rho}}{\frac {df(\rho)}{d\rho}} -{\frac {\left (\Lambda+1-\xi\right )}{\rho}}f(\rho)=0
 \label{coulombaim}
\end{equation}
which is now amenable to an AIM solution. In order to find the exact
energy eigenvalues, we define $\lambda_0$ and $s_0$  as
\begin{equation}
\lambda_0=-{\frac {\left (\rho-2 \Lambda-2\right )}{\rho}},
\hspace{0.5cm} s_0=-{\frac {\left (\Lambda+1-\xi\right )}{\rho}}
\label{ls}
\end{equation}
Using the quantization condition given by equation
(\ref{kuantization}), the $\xi$ values take the form

\begin{equation}
\xi_{1}=\Lambda +1,  \hspace{0.5cm} \xi_{2}=\Lambda+2,
\hspace{0.5cm} \xi_{3}=\Lambda+3,\hspace{0.5cm} \ldots
\end{equation}
which can be generalized as
\begin{equation}
\xi_{n}=\Lambda +n^{'} 
\end{equation}
Inserting $\xi$ and $\Lambda$ in equation (\ref{kýsaltmalar}) and
defining the principal quantum number as $n=n^{'}+J$, we obtain the
exact bound state eigen-energies:
\begin{equation}
E_{nJ}=mc^2\left[1+\frac{(\alpha Z)^{2}}{\left(n-J-\frac{1}{2}+\sqrt
{(J+\frac{1}{2})^2-(\alpha Z)^{2}}\right)^{2}}
\right]^{-\frac{1}{2}}
\end{equation}

which is in agreement with the results of the references
\cite{11,12} for the same potential. The binding energy $B_{nJ}$ can
be calculated from $B_{nJ}=mc^2-E_{nJ}$.

We can also construct the corresponding eigenfunctions using AIM as
\begin{equation}
f_n (\rho) = \left( { - 1} \right)^nC_2(\sigma)_n~{ }_1F_1 \left(
{ - n,\sigma ;\rho} \right)
\end{equation}
which gives
\begin{equation}
\label{Fr}
F(\rho)={\rho}^{\Lambda+1}{e^{-\frac{1}{2}\rho}}\left[\left( { - 1}
\right)^nC_2(\sigma)_n~{ }_1F_1 \left( { - n,\sigma ;\rho}
\right)\right]
\end{equation}
where
\begin{equation}
\sigma = 2\Lambda + 2 \hspace{0.5cm} \mbox{and} \hspace{0.5cm}
\left( \sigma \right)_n = \frac{\Gamma \left( {\sigma + n}
\right)}{\Gamma \left( \sigma \right)}
\end{equation}
Other components of the the wave functions ($G(\rho)$, $H_{1}(\rho)$
and $H_{ - 1}(\rho)$) can be obtained through equations (\ref{c1})
to (\ref{c2}) using $F(\rho)$.
\section{Anharmonic Oscillator}
\label{anharmonic} In this section, we present the application of
the Asymptotic Iteration Method (AIM) to non-trivial problems. We
have thus chosen a vector potential of type:
\begin{equation}
U_V=r^{2\xi}, \quad \xi=2, 3 \ldots
\end{equation}
Taking $\xi=2$, the second-order DKP equation becomes as follows
\begin{equation}\label{dkp2k}
{\frac {d^{2}}{d{r}^{2}}}F(r)+\left ({\frac
{{E}^{2}-2\,E{r}^{4}+{r}^{
8}-{m}^{2}{c}^{4}}{{h}^{2}{c}^{2}}}-{\frac {J\left (J+1\right
)}{{r}^{ 2}}}\right )F(r)=0
\end{equation}
In order to solve this equation with AIM, we propose the following
wave function to transform it to an equation similar to
equation~(\ref{diff}):
\begin{equation}
F(r)={e^{-1/2\,\beta\,{r}^{2}}}f(r)
\end{equation}
where $\beta$ is an arbitrarily introduced constant to improve the
convergence speed of the method. We take it $\beta$=5 as in
reference \cite{20} to compare with their non-relativistic results
for a similar problem. By taking $\hbar=c=m=1$ and $J=0$ (s-state)
for simplicity and inserting this wave function into
equation~(\ref{dkp2k}), we obtain
\begin{equation}\label{pert}
{\frac {d^{2}}{d{r} ^{2}}}f(r)=\left (-{E}^{2}+ 2\,E{r}^{4}+
 \beta+1-{\beta}^{2}{r}^{2}-{r}^{8}\right )f(r)+ 2\,\beta\,r{\frac
{d}{dr}}f(r)
\end{equation}
which can be now solved by AIM. Here, the $s_0(r)$ and
$\lambda_0(r)$ are as follows
\begin{equation} \label {s0l0}
s_0(r)=\left (-{E}^{2}+ 2\,E{r}^{4}+
 \beta+1-{\beta}^{2}{r}^{2}-{r}^{8}\right ), \quad
 \lambda_0(r)= 2\,\beta\,r
\end{equation}
In order to obtain the energy eigenvalues from
equation~(\ref{pert}), using equation~(\ref{iter}), we obtain the
$s_k(r)$ and $\lambda_k(r)$ in terms of $s_0(r)$ and $\lambda_0(r)$.
Then, using the quantization condition of the method given by
equation \ref{kuantization}, we obtain the energy eigenvalues. This
straightforward application of AIM gives us the energy eigenvalues,
however, we have observed that the energy eigenvalues oscillate and
do not converge within a reasonable number of iteration. The
sequence appears to converge when the number of iterations $k
\leq\simeq$ 30, but then it begins to oscillate as the iteration
number $k$ increases. This result violates the principle behind the
AIM; as the number of iteration increases, the method should
converge and should not oscillate. We have noticed that the first
reason for the oscillatory behavior is the $r^8$ term and the second
but less serious reason is the $E^2$ term.

Therefore, in order to overcome this problem, we have used a
perturbation approach within the framework of the AIM, similar to
reference \cite{21}. In order to apply the perturbation, we
introduce a parameter $\gamma$ for $s_0(r)$ in
equation~(\ref{s0l0}):
\begin{equation} \label {s0pert}
s_0(r)=\left (-{E}^{2}+ 2\,E{r}^{4}+
 \beta+1+\gamma (-{\beta}^{2}{r}^{2}-{r}^{8})\right )
\end{equation}
$\gamma$ is an artificially introduced perturbation expansion
parameter and at the end of the calculations, it will be seen that
it is equal to 1. After this, equation~(\ref{kuantization}) becomes
\begin{equation}\label{kuant-per}
  \delta_{k}(x,\gamma)=\lambda_{k+1}(x,\gamma)s_{k}(x,\gamma)-\lambda_{k}(x,\gamma)s_{k+1}(x,\gamma)=0
\end{equation}
If we expand $\delta(x,\gamma)$ near $\gamma$=0, we obtain the
following series:
\begin{equation} \label{delta-exp}
\delta_{k}(x,\gamma)=\delta_{k}(x,0)+{\frac {\gamma}{1!}}{\frac
{\partial\delta_{k}(x,\gamma)}{\partial\gamma}}\Big|_{\gamma=0}
+{\frac{\gamma^{2}}{2!}}{\frac{\partial^{2}\delta_{k}(x,\gamma)}{\partial\gamma^{2}}}\Big|_{\gamma=0}
+{\frac{\gamma^{3}}{3!}}{\frac{\partial^{3}\delta_{k}(x,\gamma)}{\partial\gamma^{3}}}\Big|_{\gamma=0}
+\ldots
\end{equation}
According to AIM, the quantization condition $\delta_{k}(x,\gamma)$
must be equal to zero:
\begin{equation}\label{delta-terms}
\delta_{k}^{(j)}(x,\gamma)={\frac{\gamma^{j}}{j!}}{\frac{\partial^{j}\delta_{k}(x,\gamma)}{\partial\gamma^{j}}}\Big|_{\gamma=0}
, \quad j=0, 1, 2, \ldots
\end{equation}
It is also suitable to expand the energy eigenvalue $E$,
\begin{equation} \label {E-exp}
E_{n}=E_{n}^{0}+ \gamma E_{n}^{1}+\gamma^{2}E_{n}^{2}+\gamma^{3}E_{n}^{3}+\gamma^{4}E_{n}^{4}+...
\end{equation}
$E_{n}$ expansion terms can be obtained by comparing the terms with
the same order of $\gamma$ in equations~(\ref{delta-terms}) and
(\ref{E-exp}). Hence, it is clear that the roots of
$\delta_{k}^{(0)}(x,0)$=0 give us the main contribution energy terms
$E_{n}^{0}$ and the roots of $\delta_{k}^{(1)}(x,0)$=0 give us the
first correction $E_{n}^{1}$ and so on.

After we apply this perturbation approach, we have obtained the
ground and the first even excited state energy eigenvalues. The
results are presented in Tables \ref{E0s} and \ref{E2s} respectively
for the ground and the first even excited state eigenvalues. In the
first column of Table \ref{E0s}, we present the $E_{0}^{0}$
(unperturbed), the second column $E_{0}^{1}$ which is the first
correction and so on. We have used the perturbation up to 5$^{th}$
term, but one can use higher terms to improve the results. However,
the effect becomes smaller as it can be seen from tables. In the
last column of Table \ref{E0s}, we show the non-relativistic results
of the Fernandez \cite{20} for the same potential to compare with
our results. For these calculations, we have observed that the first
term ($E_{0}^{0}$) in the expansion (\ref{E-exp}) converges around
$k$=30 iterations, however, the correction terms require higher
iterations and start to converge around $k$=50.

In Table \ref{E2s}, we show the first even excited state energy
eigenvalues. Again, the perturbation is calculated up to 5$^{th}$
and the first term ($E_{0}^{0}$) converges around $k$=35 iterations,
however, the correction terms require higher iterations and we have
run them up to $k$=50 iteration.

\section{Conclusion}
This paper has presented a different approach, the asymptotic
iteration method, to the calculation of the non-zero angular
momentum solutions of the relativistic Duffin-Kemmer-Petiau
equation. Exact eigenvalues and eigenfunction for the relativistic
Duffin-Kemmer-Petiau oscillator and Coulomb problems are derived
easily. The advantage of the asymptotic iteration method is that it
gives the eigenvalues directly by transforming the second-order
differential equation into a form of ${y}''$ =$ \lambda _0 (r){y}' +
s_0 (r)y$. The exact wave functions are easily constructed by
iterating the values of $s_0$ and $\lambda_0$. We have also shown
how to solve the non-trivial problems with the help of the
perturbation theory within the framework of the asymptotic iteration
method. The method presented in this study is general and worth
extending to the solution of other interaction problems.
\section*{Acknowledgments}
This work is supported by the Turkish Science and Research Council
(T\"{U}B\.{I}TAK), Grant No: TBAG-2398 and Erciyes
University-Institute of Science: Grant no: FBA-03-27, FBT-04-15,
FBT-04-16. Authors would like to thank Professors Y. Nedjadi, F. M.
Fernandez and Dr. H. \c{C}ift\c{c}i for useful comments, providing
some materials and reading the manuscript.

\newpage
\begin{table}[htbp]
\caption{Ground state energy of the anharmonic oscillator where $k$
is the iteration number ($n=0$, $\hbar=c=m=1$ and $J=0$ (s-state))}
\begin{center}
\begin{tabular}{ccccccccc}
\hline\hline
 $k$&$E_{0}^{0}$&$E_{0}^{1}$ &$E_{0}^{2}$ &$E_{0}^{3}$&$E_{0}^{4}$&$E_{0}^{5}$&$E_{0}$&$E_{0}$ \cite{20}\\
\hline
 5& 2.478891& -0.481521& -0.171317& -0.087565& -0.038408& -0.030214& 1.669866&             \\
10& 2.477792& -0.485884& -0.158642& -0.080739& -0.054255& -0.036055& 1.662217& 1.325073435 \\
15& 2.477837& -0.485459& -0.159187& -0.082888& -0.052218& -0.035973& 1.662112& 1.147766154 \\
20& 2.477839& -0.485450& -0.159249& -0.083021& -0.051830& -0.035991& 1.662298& 1.072223000 \\
25& 2.477838& -0.485452& -0.159247& -0.082987& -0.051875& -0.036052& 1.662225& 1.062711298 \\
30& "       & "        & -0.159246& -0.082983& -0.051885& -0.036069& 1.662203& 1.060482716 \\
35& "       & "        & -0.159246& -0.082984& -0.051885& -0.036062& 1.662209& 1.060372025 \\
40& "       & "        &         "& "        & -0.051884& -0.036060& 1.662212& 1.060362059 \\
45& "       & "        &         "& "        & "& -0.036061& 1.662211& 1.060362077 \\
50& "       & "        &         "& "        & "& "& "& 1.060362091 \\
55& "       & "        &         "& "        & "& "& "& 1.060362091 \\
60& "       & "        &         "& "        & "& "& "& 1.060362090 \\
65& "       & "        &         "& "        & "& "& "& " \\
70& "       & "        &         "& "        & "& "& "& " \\
\hline\hline
\end{tabular}
\label{E0s}
\end{center}
\end{table}

\begin{table}[htbp]
\caption{First even excited state energy of the anharmonic
oscillator where $k$ is the iteration number ($n=2$, $\hbar=c=m=1$
and $J=0$ (s-state))}
\begin{center}
\begin{tabular}{cccccccc}
\hline\hline
 $k$&$E_{2}^{0}$&$E_{2}^{1}$ &$E_{2}^{2}$ &$E_{2}^{3}$&$E_{2}^{4}$&$E_{2}^{5}$&$E_{2}$\\
\hline
 5& 5.698344& -1.478344&  0.027555& -2.574579& -1.656472&  3.568314& 3.584818 \\
10& 5.370588& -0.911718& -0.040349& -0.433823& -0.040637& -0.068883& 3.875178 \\
15& 5.413951& -0.974055& -0.267240& -0.202274& -0.037847& -0.001956& 3.930579 \\
20& 5.415995& -0.992837& -0.286535& -0.104029& -0.066451& -0.078990& 3.887153 \\
25& 5.415458& -0.991076& -0.277803& -0.121674& -0.072980& -0.054753& 3.897172 \\
30& 5.415453& -0.990550& -0.276740& -0.127545& -0.073126& -0.040145& 3.907347 \\
35& 5.415460& -0.990604& -0.277153& -0.126693& -0.071566& -0.043842& 3.905602 \\
40& 5.415460& -0.990626& -0.277226& -0.126297& -0.071080& -0.046266& 3.903965 \\
45& 5.415460& -0.990623& -0.277201& -0.126333& -0.071315& -0.045760& 3.904228 \\
50& 5.415460& -0.990622& -0.277194& -0.126358& -0.071409& -0.045369& 3.904508 \\
\hline\hline
\end{tabular}
\label{E2s}
\end{center}
\end{table}

\end{document}